\RequirePackage{fix-cm}
\documentclass[twocolumn]{svjour3}
\smartqed
\usepackage{graphicx}
\usepackage{mathpazo}
\usepackage[mathpazo]{flexisym}
\usepackage{breqn}
\usepackage[noadjust]{cite}
\usepackage{amssymb}
\usepackage{array}
\usepackage{stfloats}
\usepackage{floatrow}
\usepackage{tabularx}
\usepackage{multirow}
\usepackage{hhline}
\usepackage[export]{adjustbox}
\usepackage{pgfplots}
\usepackage[T1]{fontenc}
\usepackage[latin9]{inputenc}
\usepackage{epsfig}
\usepackage{collcell}
\usepackage{pgf}
\pgfplotsset{compat=1.14}
\usepackage{tabularx}
\newcolumntype{C}[1]{>{\centering\arraybackslash}p{#1}}
\usepackage{caption}

\begin{document}
\sloppy

\title{RemNet: Remnant Convolutional Neural Network for Camera Model Identification}

\author{Abdul Muntakim Rafi$^{1}$ \and Thamidul Islam Tonmoy$^2$ \and Uday Kamal$^3$ \and Q. M. Jonathan Wu$^1$ \and Md. Kamrul Hasan$^{3*}$}

\institute{Abdul Muntakim Rafi \at 
            \email{rafi11@uwindsor.ca}
           \and
           Thamidul Islam Tonmoy \at
              \email{ttonm001@ucr.edu}
            \and
            Uday Kamal \at \email{udday2014@gmail.com}           
           \and
            Q. M. Jonathan Wu \at
            \email{jwu@uwindsor.ca}
           \and
          $^{*}$Md. Kamrul Hasan \at  \email{khasan@eee.buet.ac.bd}
          \and 
          $^{1}$Department of Electrical and Computer Engineering, University of Windsor, Windsor, Canada\and 
          $^2$Department of Bioengineering, University of California, Riverside, USA\and
          $^3$Department of Electrical and Electronic Engineering, Bangladesh University of Engineering and Technology, Bangladesh.
           \at
           $^{*}$Corresponding author.
}

\date{Received: date / Accepted: date}

\maketitle

\begin{abstract}
Camera model identification (CMI) has gained significant importance in image forensics as digitally altered images are becoming increasingly commonplace. In this paper, a novel convolutional neural network (CNN) architecture is proposed for CMI with emphasis given on the preprocessing task considered to be inevitable for removing the scene content that heavily obscures the camera model fingerprints. Unlike the conventional approaches where fixed filters are used for preprocessing, the proposed remnant blocks, when coupled with a classification block and trained end-to-end minimizing the classification loss, learn to suppress the unnecessary image contents dynamically. This helps the classification block extract more robust camera model-specific features for CMI from the remnant of the image. The whole network, called RemNet, consisting of a preprocessing block and a shallow classification block, when trained on 18 models from the Dresden database, shows 100\% accuracy for 16 camera  models with an overall accuracy of 97.59\% on test images from unseen devices, outperforming the state of the art deep CNNs used in CMI. Furthermore, the proposed remnant blocks, when cascaded with the existing deep CNNs, e.g., ResNet, DenseNet, boost their performances by a large  margin. The proposed approach proves to be very robust in identifying the source camera models, even if the original images are post-processed. It also achieves an overall accuracy of 95.11\% on the IEEE Signal Processing Cup 2018 dataset, which indicates its generalizability.

\keywords{Digital Image Forensics \and Camera Model Identification \and Convolutional Neural Networks \and Remnant Block}

\end{abstract}

\section{Introduction}
\label{sec_introduction}
Camera  model  identification  (CMI)  has gained significant momentum in recent years for information forensics as digitally altered images  are  becoming  more  pervasive  in  electronic  media \cite{stamm2013information}. The increased usage of digital images in our everyday-life for entertainment, social networking, and more importantly in legal and security issues is, therefore, raising authenticity concern regarding the source of an image and its content, especially  when  presented  to  a  court  as  an  evidence \cite{san2006source}. Furthermore, the available professional image editing tools, though intended for entertainment purposes, are also facilitating image manipulation for illegal acts, making the problem of CMI even crucial.  
Although the metadata of an image contains some information about the source, it is not a reliable metric to determine the source since this data can be forged \cite{stamm2013information}. Besides, the metadata of the digital images are mostly unavailable when shared online in social media. Moreover, while sharing online, images go through various post-processing operations which destroy the trace of the source information to some extent, making the identification even more difficult \cite{castiglione2013experimentations}. As a result, the task of identifying the camera model is continually becoming more challenging. Therefore, a forensic analyst has to resort to image processing and analysis techniques to identify the camera model with which an image was taken.

A number of methods have been proposed in the literature for blind identification of a camera model. An extensive review of these methods can be found in \cite{stamm2013information, kirchner2015forensic}. Initially, researchers have tried to merge external features, e.g., watermarks, device-specific-code, etc., present in an image for device identification \cite{piva2013overview}. However, adding different extrinsic features to every single camera being used has proved to be an unmanageable task \cite{farid2009image}. As a result, the focus has shifted towards detecting intrinsic camera features, such as the color filter array (CFA) pattern \cite{bayram2005source}, interpolation algorithms and image quality metrics (IQM) \cite{kharrazi2004blind, gloe2012feature}. Device-specific camera detection schemes have also been proposed, where noise patterns like the photo response non-uniformity (PRNU) have been exploited to identify the device \cite{dirik2007source, fridrich2006digital, filler2008using}. At the same time, forensic researchers have developed device invariant CMI algorithms \cite{thai2014camera, lukas2006digital}. Most of these methods attempt to estimate the model-specific artifacts that are introduced into an image during the image capturing process \cite{cao2009accurate}. In this approach, the second-order statistics of the CFA pattern \cite{swaminathan2007nonintrusive} and 3D co-occurrence matrices \cite{chen2015camera, marra2017study} have been used as feature vectors to successfully detect camera models with state of the art accuracy.

Recently, researchers have adopted data-driven approaches and made efforts to solve the CMI problem using convolutional neural networks (CNNs). A common practice while using convolutional neural networks (CNNs) in digital image forensics is to perform some preprocessing on the input images to refrain the network from learning features related to the image content, at the same time, associate them to learn the camera models specific contents. Conventional median or high-pass filter has been used in some works prior to feeding images into CNNs \cite{chen2015median,tuama2016camera}. However, the reason behind using these fixed kernel coefficients or a particular kernel size is not well explained in those works, thereby requiring human intervention in designing these filters. What is more, these filters may not generalize on different datasets. In addition, as mentioned in \cite{lukas2006digital}, the signatures left by different components of the image acquisition pipeline have different frequency ranges because demosaicing and vignetting leave low-frequency patterns whereas the SPN introduces high-frequency components. Therefore, using a fixed high-pass filter may result in a loss of valuable camera model specific features. Similarly, a specific kernel size for median filtering may not serve the purpose optimally. To overcome such difficulties with the conventional fixed filters, Bayar and Stamm \cite{bayar2017design} have proposed a data-driven constrained convolutional layer which is shown to be superior in performance to both median and high-pass filters. However, the constrained convolution, originally proposed for image manipulation detection in \cite{bayar2016deep}, that extracts prediction error features in the preprocessing stage does not explain how these features retain camera model-specific features. Nevertheless, the improvement of results associated with the different preprocessing schemes has made it very clear that a customized preprocessing operation should be explored thoroughly in this field.

More recently, another category of works has emerged that does not employ any preprocessing stage that has been shown to facilitate feature extraction for CMI \cite{yang2017source, bondi2017first, yao2018robust, rafi2019application}. In \cite{yang2017source}, a concept of fusion residual network (FRN) is proposed which uses the idea of using multi-scale receptive fields on an input image. The FRN extracts intrinsic features from the input image through convolutions with different kernel sizes and then concatenates the extracted feature maps. Bondi et al. \cite{bondi2017first} have proposed a combination of CNN and support vector machine (SVM) to classify camera models, where they have used CNN to extract camera-specific artifacts. On the other hand, Yao et al. \cite{yao2018robust} have proposed a comparatively deeper CNN architecture for CMI. Although these approaches show promising performances, none of the authors have investigated if the performance of their networks can be ameliorated with the incorporation of dynamic preprocessing filters. In \cite{rafi2019application}, the authors explore the
performance of DenseNet \cite{huang2017densely} using three different image scales (64 X 64, 128 X 128, 256 X 256), which they find beneficial to the CNN model.

Despite the breadth of works performed in this field, little attention has been given to the identification of camera models from images of unseen devices-- the devices whose images have not been used in training the neural networks. Kirchner and Gloe have emphasized on this issue by proposing an evaluation criterion that uses disjoint subsets of devices for training and testing CMI methods to replicate real-world scenarios \cite{kirchner2015forensic}. Moreover, the performance of the existing CMI methods on post-processed images, e.g., JPEG compressed, resized, gamma-corrected, etc., is not well studied. Although some researchers have explored the case of detecting image manipulation discretely \cite{li2008detecting, stamm2010forensic, kee2011digital, bayar2016deep, bayar2017design}, not many have tried to identify source camera model from post-processed images. In reality, a robust CMI network needs to correctly predict the source camera model of an image that may have gone through diverse \emph{post-processing} and captured by an \emph{unseen} device.

In this work, we propose RemNet, a novel CNN-based architecture to perform CMI task on extensively post-processed images acquired from unseen devices. The major constituent part of RemNet is the data-adaptive preprocessor that comprises of several remnant blocks. Unlike the conventional fixed-filter based approaches, our preprocessor can dynamically adapt its parameters to perform required preprocessing task suitable for the subsequent classification block. We also adopt a modular structure in our architecture which enables any CNN-based classifier to be cascaded with our proposed preprocessor.  Our extensive experimentation shows that RemNet not only surpasses the state of the art CMI networks but also enhances their performance if used in cascade with our proposed preprocessor.

The rest of the paper is organized as follows. Section \ref{sec_proposed_method} presents a detailed description of our proposed method, along with the motivations and intuitions behind designing it. Section \ref{sec_experimental_results} provides a thorough discussion of the training and evaluation procedure, along with the experimental results obtained after testing the model with different datasets. Finally, we conclude in Section \ref{sec_conclusion}.

\begin{figure*}[!t]
	\centering
	\includegraphics[width = 6.5in]{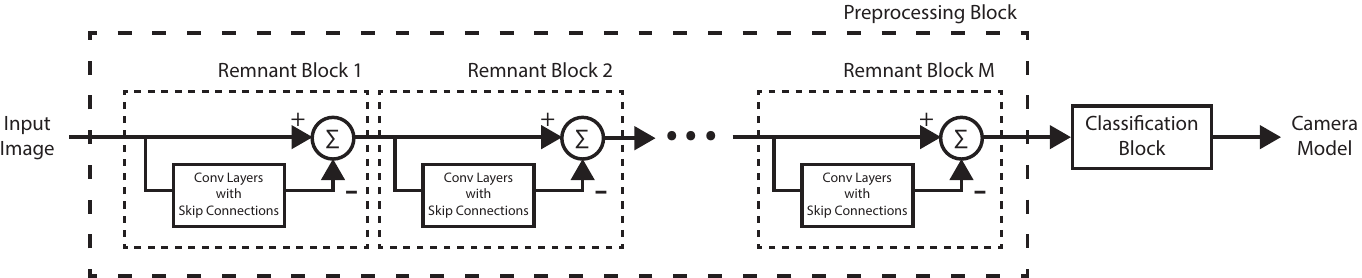}
	\caption{Block diagram of our proposed RemNet.}
	\label{fig_pipeline}
\end{figure*}

\section{Proposed CNN Model}
\label{sec_proposed_method}
In designing CNNs for image forensic tasks, it has been a common practice to use a preprocessing scheme to suppress the image contents and intensify the minute signatures induced by the image acquisition pipeline \cite{chen2015median, tuama2016camera, bayar2017design}. However, the methods reported so far suffer from their own drawbacks of using either fixed kernels or constraints as described earlier. The main objective of this work is, therefore, to introduce a preprocessing scheme that is completely data-driven but without any imposed constraints or fixed kernels. To this end, we design a novel CNN architecture called RemNet. 

\begin{figure*}[!t]
	\centering
	\includegraphics[width = 6.5in]{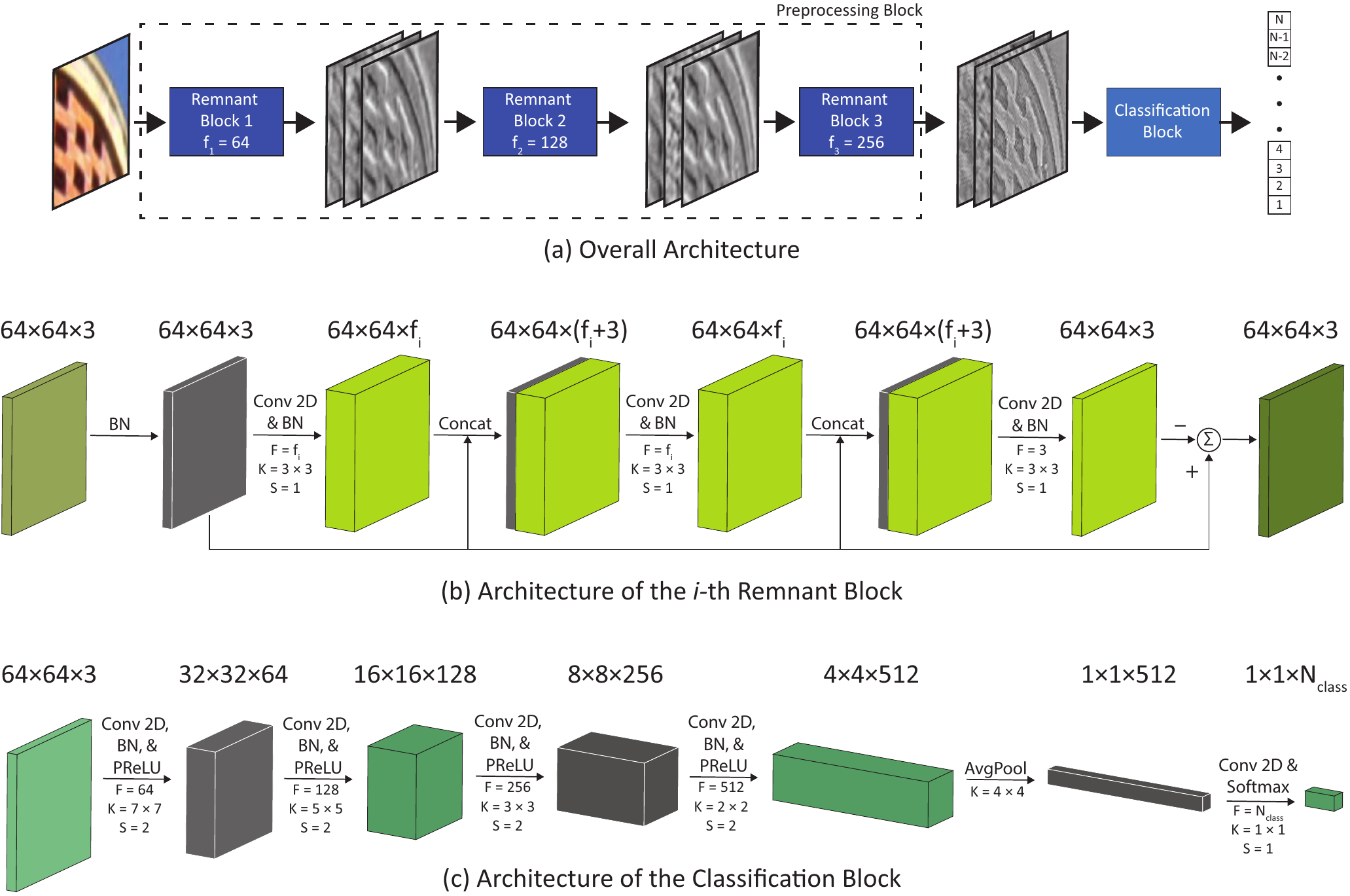}
	\caption{The architecture of our proposed RemNet. (a) Illustrates the overall architecture with three remnant blocks with one classification block. The architectures of the remnant and classification blocks are depicted in (b) and (c), respectively. In (b) and (c), AvgPool, BN, and Conv2D represent average pooling, batch normalization, and 2D convolution, respectively. The letters F, K, and S represent the number of filters, their kernel sizes, and strides, respectively, in the corresponding convolution layers. The letter $\rm N_{class}$ represents the number of camera models.}
	\label{fig_network}
\end{figure*}

RemNet is comprised of two major building blocks-- a data-driven preprocessing block used at the beginning of the network which is followed by a classification block (see Fig. \ref{fig_pipeline}). These blocks are trained end-to-end so that the preprocessing block acts as a data-driven custom preprocessing scheme on the input image that learns to suppress image contents to some extent as required for better minimization of the loss function and intensifies camera model-specific feature-rich portions of the image at its output. The details of our proposed network architecture are presented in the following.
	
\subsection{Preprocessing Block}
\label{sec_preprocessing_block}
The preprocessing block consists of several remnant blocks. The detailed architecture of the remnant block is shown in Fig. \ref{fig_network}(b). Each block consists of 3 convolutional layers with kernel size $3 \times 3$ followed by BN. Inside each block, the feature space is widened from $64 \times 64 \times 3$ to $64 \times 64 \times f_i$ in the first 2 convolutional layers and then reduced to $64 \times 64 \times 3$ again in the last convolutional layer. The choices for $f_i$ in the consecutive remnant blocks are 64, 128, and 256, respectively. Finally, to generate the residue, the output of the final convolutional layer in a block is subtracted from the input in a pixel-wise manner. As the convolutional layers are followed by batch normalization (BN) layer, in spite of directly using the input, we use the batch normalized version of it. Our intuition behind such architectural choice is to enable a remnant block to learn the required transformation that would disintegrate the undesired contents so that the subsequent subtraction operation can suppress them and generate forensic feature enriched residue. But there is still a possibility of losing some important forensic information after such intermediate convolution operations. As the subsequent blocks operate on the residue generated by the previous block, such information loss would gradually build up, causing considerable degradation of the model's performance. The input information must be preserved as much as possible throughout the block to alleviate this problem. In order to ensure this, we include several skip connections so that the input to a remnant block is propagated to every convolutional layer inside that block. Even if some of the minute features are lost in a layer, it is regenerated through the skip connections (see Fig. \ref{fig_network}(b)). This also prevents the vanishing of gradient-flow during training. We do not use any activation function in our remnant blocks because we prefer to build the remnant blocks as linear filters that will act as optimal preprocessors for CMI. The contribution of the remnant blocks is experimentally verified in our experimental results section (see Table \ref{tab_design}). 
	
There are several hyperparameter choices in the final structure of our preprocessing scheme: the number of remnant blocks, the depth of a single block, the number of filters in each layer, and kernel size-- all of these are set using cross-validation.
	
The remnant blocks are somewhat influenced by the highway networks proposed by Srivastava et al. in \cite{srivastava2015highway}. A plain convolutional layer applies a linear transformation $H$ (parameterized by $\mathbf{W}_{\mathbf{H}}$) on its input $\mathbf{x}$ to produce its output $\mathbf{y}$:
\begin {equation}
	\mathbf{y}=H\left(\mathbf{x}, \mathbf{W}_{\mathbf{H}}\right),
\end {equation}
where $H$ is usually an affine transformation followed by a nonlinear activation function, but it may take different forms for different tasks.

\begin{figure*}[!t]
	\centering
	\includegraphics[width = 6.5in]{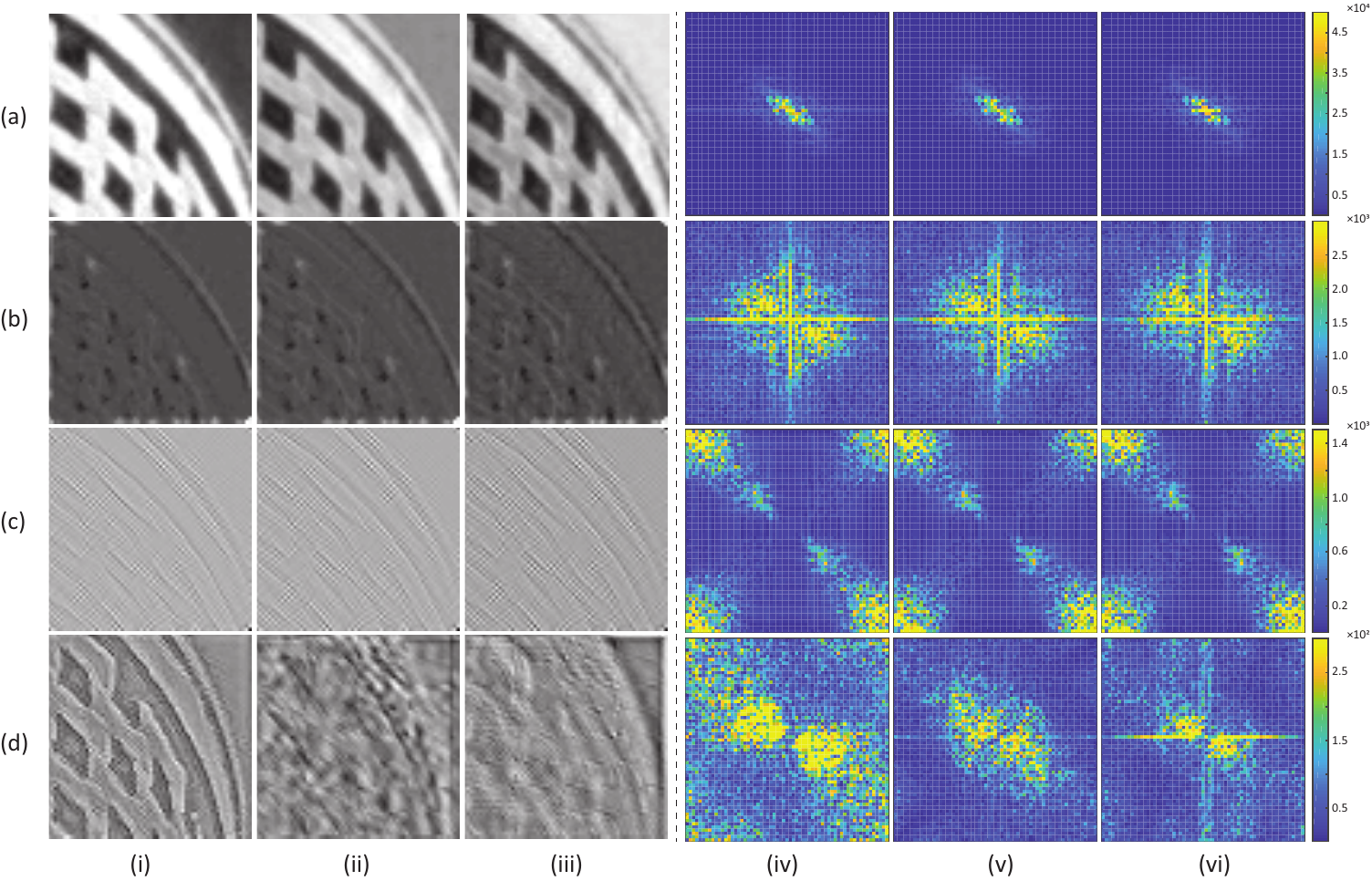}
	\caption{Comparison of outputs of various preprocessing schemes. (a) Input image, (b) median filter residue, (c) high-pass filter output, and (d) output of the third remnant block of our proposed RemNet. Columns (i), (ii), and (iii) correspond to different output channels, whereas columns (iv), (v), and (vi) depict their frequency responses, respectively.}
	\label{fig:output}
\end{figure*}
	
For a highway network, two nonlinear transforms $T(\mathbf{x}, \mathbf{W}_{\mathbf{T}})$ and $C(\mathbf{x}, \mathbf{W}_{\mathbf{C}})$ are defined such that
\begin {equation}
	\label{highway}
	\mathbf{y}=H\left(\mathbf{x}, \mathbf{W}_{\mathbf{H}}\right) \cdot T\left(\mathbf{x}, \mathbf{W}_{\mathbf{T}}\right)+\mathbf{x} \cdot C\left(\mathbf{x}, \mathbf{W}_{\mathbf{C}}\right),
\end {equation}
where $T$ is the transform gate and $C$ is the carry gate. $T$ controls how much of the activation is passed through and $C$ controls how much of the unmodified input is passed through. Our remnant blocks are motivated by these two gating units. We make significant modifications in our transformation function $H$ because of the nature of the operation we want to perform. As the remnant blocks are intended to be designed as a linear preprocessor, as stated before, we avoid the use of nonlinear activation functions. Also, we make use of multiple intra-block skip connections in our remnant block to preserve input information throughout a block. We use a pixel-wise subtraction operation that regulates the flow of information and alleviates the loss of information through successive convolutional operations. For the above-mentioned reasons, our transform and carry gate are linear in nature and we set $T$ and $C$ as $-$1 and 1, respectively. As a result, (\ref{highway}) becomes
\begin {equation}
	\mathbf{y}=\mathbf{x} -H\left(\mathbf{x}, \mathbf{W}_{\mathbf{H}}\right).
\end {equation}
	
The residual network (ResNet) \cite{he2016deep} is also a variant of the highway network \cite{greff2016highway} where the choices for both $T$ and $C$ are 1 for the residual blocks. However, the transformation $H$ used in \cite{he2016deep} works as a nonlinear feature extractor whereas the $H$ of our remnant blocks performs linear filtering operation. In addition, ResNet does not use any skip connections. 

To demonstrate that the dynamically designed remnant blocks truly performs the desired preprocessing task, we show in Fig. \ref{fig:output} the outputs of the final remnant block along with their frequency characteristics for a randomly selected image. We also make a spatial and frequency domain comparison of the conventional filters, e.g., median and high-pass filters used in \cite{chen2015median, tuama2016camera}, respectively. Fig. \ref{fig:output}(a) shows the RGB image, Figs. \ref{fig:output}(b)-\ref{fig:output}(d) show the median filtered residue, high-pass filtered output, and the output of the last remnant block, respectively. If we observe the frequency domain representation of the outputs, we notice that conventional fixed filters are constrained in the frequency domain as compared to our remnant blocks since the conventional filters apply the same frequency domain transformation on all the channels equally. However, it is well known that the sensor pattern noise is not uniformly distributed throughout all three channels \cite{prnu}, and Lukas et al. \cite{lukas2006digital} have explicitly stated that both low and high-frequency information are required for CMI. We, therefore, claim that our data-adaptive preprocessing performs better filtering operation, preserving the camera signature from a wide range of frequencies, which is empirically justified by our experimental results presented in Section \ref{sec_experimental_results}.
	
\subsection{Classification Block}
\label{sec_classification_block}
The output of the final remnant block, of size $64\times64\times3$, is passed to a classification block which is outlined in Table \ref{tab_arch}. The aim of this module is to extract higher-level camera model-specific features, reduce the dimensions of the feature vectors, and eventually generate a class probability of the source camera model of the input image. The classification block is trained end-to-end with the remnant blocks. Therefore, it forces the remnant blocks to suppress unnecessary contents, enhance the useful ones, and then generate a remnant of the image which contains rich camera model fingerprints for better minimization of the classification loss function.

\begin{table}[!b]
	\centering
	\caption{Architecture of our proposed RemNet}
	\begin{tabular}{C{0.9in} C{0.9in} C{0.9in}}
		\hline
		Layers & Output Size & Kernels*\\
		\hline
		\multicolumn{3}{c}{\textbf{Preprocessing Block}}\\
		\hline
		Remnant Block 1 & 64$\times$64$\times$3 & $f_1$ = 64\\
		Remnant Block 2 & 64$\times$64$\times$3 & $f_2$ = 128\\
		Remnant Block 3 & 64$\times$64$\times$3 & $f_3$ = 256\\
		\hline
		\multicolumn{3}{c}{\textbf{Classification Block}}\\
		\hline
		Conv 2D, BN, \& PReLU & 32$\times$32$\times$64 & F = 64, K = 7$\times$7, S = 2\\
		Conv 2D, BN, \& PReLU & 16$\times$16$\times$128 & F = 128, K = 5$\times$5, S = 2\\
		Conv 2D, BN, \& PReLU & 8$\times$8$\times$256 & F = 256, K = 3$\times$3, S = 2\\
		Conv 2D, BN, \& PReLU & 4$\times$4$\times$512 & F = 512, K = 2$\times$2, S = 2\\
		Average Pool & 1$\times$1$\times$512 & K = 4$\times$4\\
		Conv 2D & 1$\times$1$\times$$N_{class}$ & F = $N_{class}$, K = 1$\times$1, S = 1\\
		Softmax & $N_{class}$ & -- \\
		\hline
		\multicolumn{3}{p{3in}}{* \footnotesize{The letters F, K, and S represent the number of filters, their kernel size, and strides, respectively, in the corresponding convolution layers. The letter $N_{class}$ represents the number of camera models.}}
	\end{tabular}
	\label{tab_arch}
\end{table}
	
The classification block has four consecutive convolution layers at the beginning. Each of the convolutional layers is followed by a BN layer and a PReLU activation. The output of the fourth convolutional layer, of size $4\times4\times512$,  is followed by an average-pooling operation, which reduces the feature vector to a size of $1\times1\times512$. Finally, we pass the average-pooled feature vector to a final convolution layer with softmax activation to generate probabilities for the $N_{class}$ number of camera models.
	
Instead of using max-pool operation, we use strided convolution to reduce the feature space in the first four convolution layers. This makes the feature reduction process learnable and much less aggressive compared to max-pool \cite{springenberg2014striving}. As per the design principles introduced in \cite{bayar2017design}, we gradually decrease the kernel size in the first convolution layers. The BN layer is included for regularization and faster convergence. 
	
Previously CNNs used the ReLU as the activation function \cite{krizhevsky2012imagenet}. But here we want to emphasize on extracting camera model fingerprints which are statistical in nature. They do not necessarily have to be positive. As we do not want to put any constraint on the feature generation, we use the PReLU activation function in our classification block. Also, when CNNs used with a PReLU activation function, it has experimentally demonstrated higher accuracy \cite{bayar2016deep}. We have also experimentally verified this in our experimental results section (see Table \ref{tab_design}).

The average-pool operation is used as per the conventional design structure of CNNs \cite{simonyan2014very, he2016deep, huang2017densely} to reduce the dimensionality of the feature space before making the final decision. We do not use fully connected layers in the classification block to keep the number of parameters lower, which in turn makes the network less prone to overfitting. This also helps the network to train faster. 

\subsection{Loss Function and Training}
\label{sec_loss_function}
 The preprocessing block consists of $M$ remnant blocks. The $i$-th remnant block performs a transformation $H_{i}$ (parameterized by $\mathbf{W}_{\mathbf{p_{i}}}$) on its input $\mathbf{x_{i}}$ (that is the output of the $(i-1)$-th remnant block) and subtracts it from its input to produce its output $\mathbf{y}_{\mathbf{p_{i}}}$:
\begin {equation}
	\mathbf{y_{\mathbf{p_{i}}}}= \mathbf{x_{i}} - H\left(\mathbf{x_{i}}, \mathbf{W}_{\mathbf{p_{i}}}\right),
\end {equation}

The output of the last remnant block, $\mathbf{y}_{\mathbf{p_{M}}}$, then becomes the input of the classifier block that applies another transformation $G$ (parameterized by $\mathbf{W}_{\mathbf{c}}$) to produce the final output $\mathbf{y}_{\mathbf{c}}$:
\begin {equation}
	\mathbf{y_{\mathbf{c}}}=G\left(\mathbf{y_{\mathbf{p_{M}}}}, \mathbf{W}_{\mathbf{c}}\right),
\end {equation}	

Finally, multiclass categorical crossentropy loss is calculated based on this output and the ground truth using the following equation:
\begin{equation}
		L = \sum_{k=1}^{N_{class}} y_{c_{i}}^{*(k)} \log \left(y_{c_{i}}^{(k)}\right),
\end{equation}
where $y_{c_i}^{*(k)}$ and $y_{c_i}^{(k)}$ are the true label and the network output of the $i$-th image at the $k$-th class among the $N_{class}$ classes, respectively,. The gradient of this loss is backpropagated to update the weights of both the preprocessing block and the classifier block of the network. Since the preprocessing block generates a residue of the input signal and the subsequent classifier will have to extract useful features from this residue alone as well as the whole network being trained in an end to end manner, the minimization of the loss function ensures that the preprocessing block learns to suppress the image contents that are irrelevant for CMI and the residue generated by it contains rich camera fingerprints.
	
\section{Experimental Results}
\label{sec_experimental_results}
To demonstrate the effectiveness of the RemNet and the remnant blocks separately, we conduct a number of experiments. In this section, we discuss those experimental results in detail. All of the experiments regarding training and implementation of the model are performed in a hardware environment which includes Intel Core-i7 8700K, 3.70 GHz CPUs and Nvidia GeForce GTX 1080 Ti (11 GB Memory) GPU. The necessary codes are written in Python and the neural network models are implemented using the Keras API (version 2.1.6) with TensorFlow-GPU (version 1.8.0) in the backend.
	
\subsection{Results on Dresden Dataset}
\label{sec_result_dresden}
We comprehensively evaluate our overall approach on the Dresden dataset \cite{dresden}. These images are captured with 73 devices of 27 different camera models. Multiple shots have been taken from several locations (e.g., office, public square, etc.) for each device. Different pictures are acquired from different viewpoints (e.g., looking on the right, on the left, etc.) for each location. We refer to different combinations of locations and viewpoints as different \emph{scenes}. The acquisition process is explained in detail in \cite{dresden}. In our work, we choose only those camera models which have more than one device so that we can keep one device separate for testing purpose. This results in discarding 8 camera models. Of the rest 19 devices, we consider two camera models, Nikon D70 and Nikon D70s, as a single model based on the work of Kirchner and Gloe. \cite{kirchner2015forensic}. Consequently, we train and test our models using the images of these 18 camera models. A brief description of the dataset used is presented in Table \ref{tab_dresden}.\\
	
\begin{table}[!b]
	\centering
	\begin{tabular}{C{0.2in} C{1in} C{0.4in} C{0.3in} C{0.3in}}
		\hline
		Serial No. & Camera Model & No. of Images & \multicolumn{2}{c}{No. of Devices} \\
		\cline{4-5}
		& & & Train and Val. & Test\\
		\hline 
		1 & Canon IXUS 70 & 363 & 2 & 1\\
		2 & Casio EX-Z150 & 692 & 4 & 1\\
		3 & FujiFilm FinePix J50 & 385 & 2 & 1\\
		4 & Kodak M1063 & 1698 & 4 & 1\\
		5 & Nikon Coolpix S710 & 695 & 4 & 1\\
		6 & Nikon D200 & 373 & 1 & 1\\
		\multirow{2}{*}{7} & Nikon D70 & \multirow{2}{*}{373} & 1 & 1\\
		& Nikon D70S & & 1 & 1\\
		8 & Olympus $\mu$1050SW & 782 & 4 & 1\\
		9 & Panasonic DMC-FZ50 & 564 & 2 & 1\\
		10 & Pentax Optio A40 & 405 & 3 & 1\\
		11 & Praktica DCZ 5.9 & 766 & 4 & 1\\
		12 & Ricoh Capilo GX100 & 559 & 4 & 1\\
		13 & Rollei RCP-7325XS & 377 & 2 & 1\\
		14 & Samsung L74wide & 441 & 2 & 1\\
		15 & Samsung NV15 & 412 & 2 & 1\\
		16 & Sony DSC-H50 & 253 & 1 & 1\\
		17 & Sony DSC-T77 & 492 & 3 & 1\\
		18 & Sony DSC-W170 & 201 & 1 & 1\\
		\hline 
		& Total & 9831 & & \\
		\hline
	\end{tabular}
	\caption{Camera models of the Dresden database used in our experiments}
	\label{tab_dresden}
\end{table}
	
\subsubsection{Training and testing strategy}
\label{sec_train_test_strategy}
Training a CMI network is challenging because of the existence of device-specific features such as PRNU noise \cite{fridrich2006digital, lukas2006digital} along with model-specific features in the image. Therefore, a network that can detect the model-specific features needs to be trained in such a way that it excludes the device-specific features as much as possible and is able to focus on the model-specific features. We solve this problem by using images from multiple devices to train our network for most camera models.
	
We first split the dataset into train, validation, and test sets in such a way that the camera device and scenes used during testing are never used for training or validation. This results in 7938, 1353 and 540 images in the train, validation and test set, respectively (see Table \ref{tab_dresden}). We refer to these sets as \emph{unaltered} train, validation, and test sets. This splitting policy, proposed in \cite{bondi2017first}, is of paramount importance so that we can be sure that the neural network does not overfit on the training data and the testing accuracy is not biased by device-specific features or the natural content of the scenes.
	
\begin{figure}[!b]
	\centering
	\includegraphics[width = 3.2in]{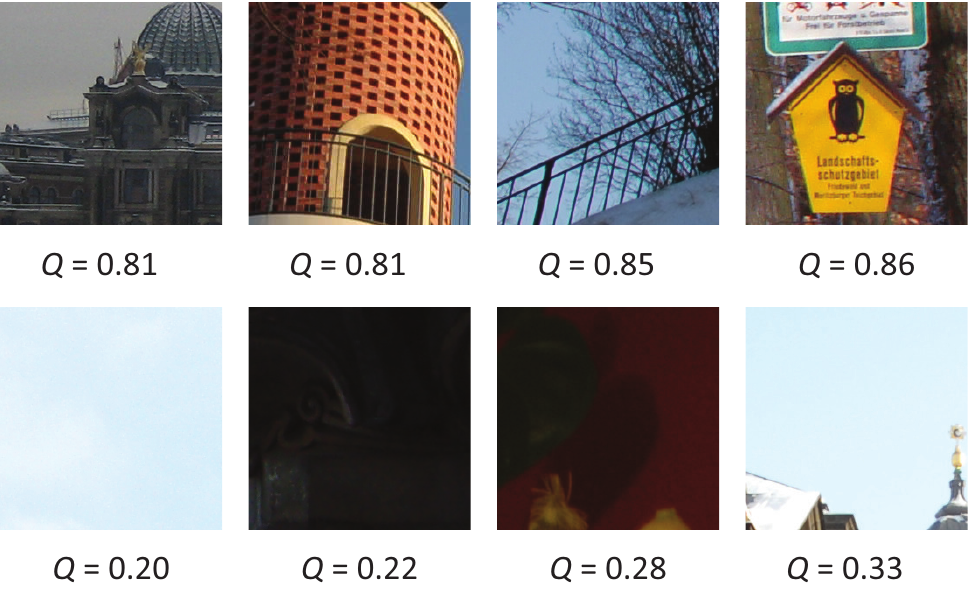}
	\caption{Examples of clusters of different qualities with their quality indices. The top row represents rich quality clusters and the bottom row represents poor quality clusters.}
	\label{fig_patch}
\end{figure}

After splitting the dataset, we extract $256\times 256$ sized clusters of pixels from the original images. However, it is to be noted that all clusters of pixels from an image are not rich in camera model-specific features. In particular, saturated and flat regions are not likely to contain enough statistical information about the camera model \cite{bondi2017first}. Therefore, different authors have used different cluster selection strategies in the literature. In \cite{yang2017source}, the authors propose a new metric to classify the image clusters into three categories: i) Smooth, ii) Saturated and iii) Others. After that, they train their network on these three categories separately and get three different networks (same architecture but different weights) on which they report the performance results for the respective categories of image clusters. On the other hand, in \cite{bondi2017first}, the authors propose a metric that gives a higher score to the image cluster with more texture, and train and test their network with these high-scoring clusters only. Since our target is to propose a single CMI network for solving the task, we need to train and test it with clusters that contain enough statistical information about the camera model. That is why we compute the quality value of a cluster as outlined in \cite{bondi2017first}. For each cluster $\mathcal{P}$ in an image, its quality $Q(\mathcal{P})$ is computed as
\begin{dmath}
	\label{eqn_quality}
	Q(\mathcal{P})=\frac{1}{3} \sum_{c\, \in[R, G, B]} \left[\alpha \cdot \beta \cdot(\mu_c - \mu_{c}^2\right) + \left(1 - \alpha) \cdot(1 - e^{\gamma \sigma_c})\right]
\end{dmath}
where $\alpha$, $\beta$, and $\gamma$ are empirically set constants (set to 0.7, 4 and $\ln (0.01)$, respectively), $\mu_c$ and $\sigma_c$, $c \in [R, G, B]$ are the mean and standard deviation of the red, green, and blue components of cluster $\mathcal{P}$, respectively. For a cluster of pixels with texture, this quality measure tends to be higher than for the overly saturated or flat clusters (see Fig.  \ref{fig_patch}). We found that this quality assessment is consistent with the `\emph{others}' category mentioned in \cite{yang2017source}. According to the definition in \cite{yang2017source}, $99.32\%$ of our high-quality clusters fall into \emph{others} category while $0.63\%$ are \emph{smooth}, and the rest $0.03\%$ are \emph{saturated}. Therefore, we can consider that our cluster selection strategy is almost identical to choosing the `others' category patches of Yang et al. \cite{yang2017source}.
	
Although we extract $256 \times 256$ sized rich quality clusters from the main image, the input patch size that we opt to use for our network is $64 \times 64$, as suggested in \cite{yang2017source, bondi2017first, yao2018robust}. During training, we select a patch of size $64 \times 64$ randomly from a cluster of $256 \times 256$ in each epoch. The idea of small input patch of $64 \times 64$  is motivated by three reasons: (i) it results in more data to train our proposed network; (ii) during the test, it enables us to generate multiple predictions for a given image and averaging over all of those predictions may ensure a more accurate classification; (iii) training our network with patches of smaller size relative to the image prevents our network from learning dominant spatial features of the image affixed directly to its contents, subsequently enabling the network to learn inherent model-specific statistical features. Also, training a network with bigger input patch size poses hardware constraints and requires more training time. 
	
Our cluster and patch selection strategy introduces statistical variations during training. The network cannot rely on seeing the same patch of size $64 \times 64$ more than once since they are randomly extracted from the $256 \times 256$  clusters in each epoch. This has a regularizing effect and forces the network to learn more robust features that generalize better across multiple samples of the input data. Our proposed cluster selection method also ensures that the input patches of $64 \times 64$ to the network are a mix of good and bad patches where good patches are dominant in number. Some of the rich quality clusters of $256 \times 256$ may contain a few bad patches of $64 \times 64$ as illustrated in Fig. \ref{fig_patch}. Therefore, during training, the network learns to extract features from saturated regions as well. This, in turn, helps our network perform well in poor quality clusters extracted from the main image, which is demonstrated in the experimental results. 
	
During training, we extract 20 rich quality clusters of size $256 \times 256$ from each image which results in 158760 and 27060 clusters for the unaltered train and validation set, respectively. We then randomly crop a $64 \times 64$ size patch from each cluster in each epoch and feed it to the network. Since we are experimenting with 18 camera models, we set $N_{class} = 18$ for our classification block. The weights of the network kernels are initialized randomly with the uniform distribution proposed by Glorot and Bengio \cite{glorot2010understanding}. We use categorical cross-entropy as the loss function and Adam \cite{kingma2014adam} as the optimizer with the exponential decay rate factors $\beta_1$ = 0.9 and $\beta_2$ = 0.999. The batch size we opt to use is 64. The initial learning rate is set to $10^{-3}$ and is decreased by a factor of $0.5$ if the validation loss does not decrease in two successive epochs. When the learning rate is reduced to $10^{-7}$, the training is stopped. In this way, we train our network for a maximum of 50 epochs and save the weight with the least validation loss for evaluation.
	
After training, we test our network on the unaltered test set comprised of $540$ images from unseen devices of 18 different camera models of the Dresden database. During testing, we select $N$ number of rich quality clusters of size $256 \times 256$ from each test image according to our quality assessment. To make a prediction for each cluster, we take the average of the predictions on all non-overlapping patches of size $64 \times 64$ it contains and assigns a camera model label $\hat{l_{j}}$ to it. The final prediction for the image is obtained through majority voting on  $\hat{l_{n}}$ for $n \in [1, N]$. In all the subsequent experiments, we use $N = 20$ unless otherwise stated. Finally, the accuracy of the network is obtained using the following equation:
\begin{equation}
	\label{eqn_acc}
	\textrm{Accuracy} = \frac{N_{corr}}{N_{tot}} \times 100 \%,
\end{equation}
where $N_{corr}$ is the number of images correctly predicted and $N_{tot}$ is the total number of images, which in this case, is $540$.

\subsubsection{Comparison of design choices}
\label{sec_design_choice}
First, we experiment with several architectural design choices of our proposed RemNet. We train and test these various designs on the unaltered dataset. The results of these experiments are presented in Table \ref{tab_design}. It is evident from the table that our proposed RemNet with 3 remnant blocks followed by a classification block with PReLU activation results in a better accuracy. The detection accuracy it achieves is 97.03\%.

\subsubsection{Comparison with state of the art networks on unaltered images}
\label{sec_result_unalt}
We compare our results with the established methods in CMI-- constrained-convolutional network \cite{bayar2017design}, fusion residual network \cite{yang2017source} and first steps toward the camera model identification with convolutional neural networks \cite{bondi2017first}. The reason behind choosing \cite{bayar2017design} and \cite{yang2017source} is that both of these works incorporate their own preprocessing scheme that agrees to our main intuition in this work. Since our rich quality clusters commensurate with the `others' category of \cite{yang2017source}, we implement the fusion residual network for the `others' category only, instead of each of the three different categories mentioned in \cite{yang2017source}. We also include \cite{bondi2017first} in our comparison as we adopt their cluster selection strategy. Recently, several works such as \cite{Chen_densenet}, \cite{barni-jpeg-cnn}, and \cite{DL_cmi} confirm the superior performance of very deep neural networks in different camera forensic applications. As a result, we also compare the performance of the RemNet with two CNN based deeper architectures namely ResNet \cite{he2016deep} and DenseNet \cite{densenet}.  For a fair comparison, we use the same input patch size, $64 \times 64$, for all the networks and the implementation of each method is made under careful scrutiny. 
	
\begin{table}[!t]
	\centering
	\begin{tabular}{ C{2.4in} C{0.5in}}
		\hline
		Design Choice & Accuracy (\%) \\
		\hline 
		Remnant Blocks + Classifier (ReLU) & 96.48 \\
		Remnant Blocks with Activation (PReLU) + Classifier (PReLU) & 96.67 \\
		\textbf{Remnant Blocks + Classifier (PReLU)} & \textbf{97.03} \\
		\hline
	\end{tabular}
	\caption{Accuracy of different design choices of RemNet trained and tested on the unaltered train and test sets of the Dresden database}
	\label{tab_design}
\end{table}
	
\begin{table}[!b]
	\begin{tabular}{C{1.4in} C{0.8in}}
		\hline
		Method & Accuracy (\%)\\
		\hline 
		Bayar and Stamm \cite{bayar2017design} & 95.56 \\
		Yang et al. \cite{yang2017source} & 94.81 \\
		Bondi et al. \cite{bondi2017first} & 90.55 \\
		ResNet \cite{he2016deep} & 92.40 \\
		DenseNet \cite{densenet} & 93.33 \\ 
		\textbf{Proposed Method} & \textbf{97.03} \\ 
		\hline
	\end{tabular}
	\caption{Accuracy of different methods trained and tested on the unaltered train and test sets of the Dresden database}
	\label{tab_result_unalt}
\end{table}
	
The results presented in Table \ref{tab_result_unalt} show that networks with preprocessing schemes perform substantially better than the other networks and our proposed RemNet outperforms all the networks with a significant margin. This observation, therefore, establishes our claim that a preprocessor is indeed necessary in CMI even for deeper architectures.\\
	
\subsubsection{Effects of Data Augmentation}
\label{sec_data_augment}
Deep neural networks have a tendency to overfit due to their large number of learnable parameters. Since these methods require a large amount of data to avoid overfitting, data augmentation is a commonly used method in training CNNs \cite{wong2016understanding}. Also, our goal is to design a robust network that can perform CMI even if the image is post-processed. To address these challenges, we use different types of post-processing methods as a form of data augmentation to increase the volume of training data. The types of augmentation that we use in this work are:
\begin{itemize}
	\item JPEG-Compression with quality factor of 70\%, 80\%, and 90\%
	\item Rescaling by a factor of 0.5, 0.8, 1.5, and 2.0
	\item Gamma-Correction using $\gamma$ = 0.8 and 1.2
\end{itemize}
We perform the aforementioned post-processing methods on the train and validation sets which increase the volume of data by 9 folds. We refer to these increased datasets as \emph{augmented} train and validation sets. The augmented datasets contain both unaltered and manipulated images.
	
\begin{table}[!b]
	\begin{tabular}{C{1.4in} C{0.8in}}
		\hline
		Method & Accuracy (\%)\\
		\hline
		Bayar and Stamm \cite{bayar2017design} & 93.89 \\
		Yang et al. \cite{yang2017source} & 95.19 \\
		Bondi et al. \cite{bondi2017first} & 92.59 \\
		ResNet \cite{he2016deep} & 95.18 \\
		DenseNet \cite{densenet} & 95.05 \\ 
		\textbf{Proposed Method} & \textbf{97.59} \\ 
		\hline
	\end{tabular}
	\caption{Accuracy of different methods trained on the augmented train set and tested on the unaltered test set of the Dresden database}
	\label{tab_result}
\end{table}
	
\begin{figure}[!b]
    \centering
    \includegraphics[width = 3.2in]{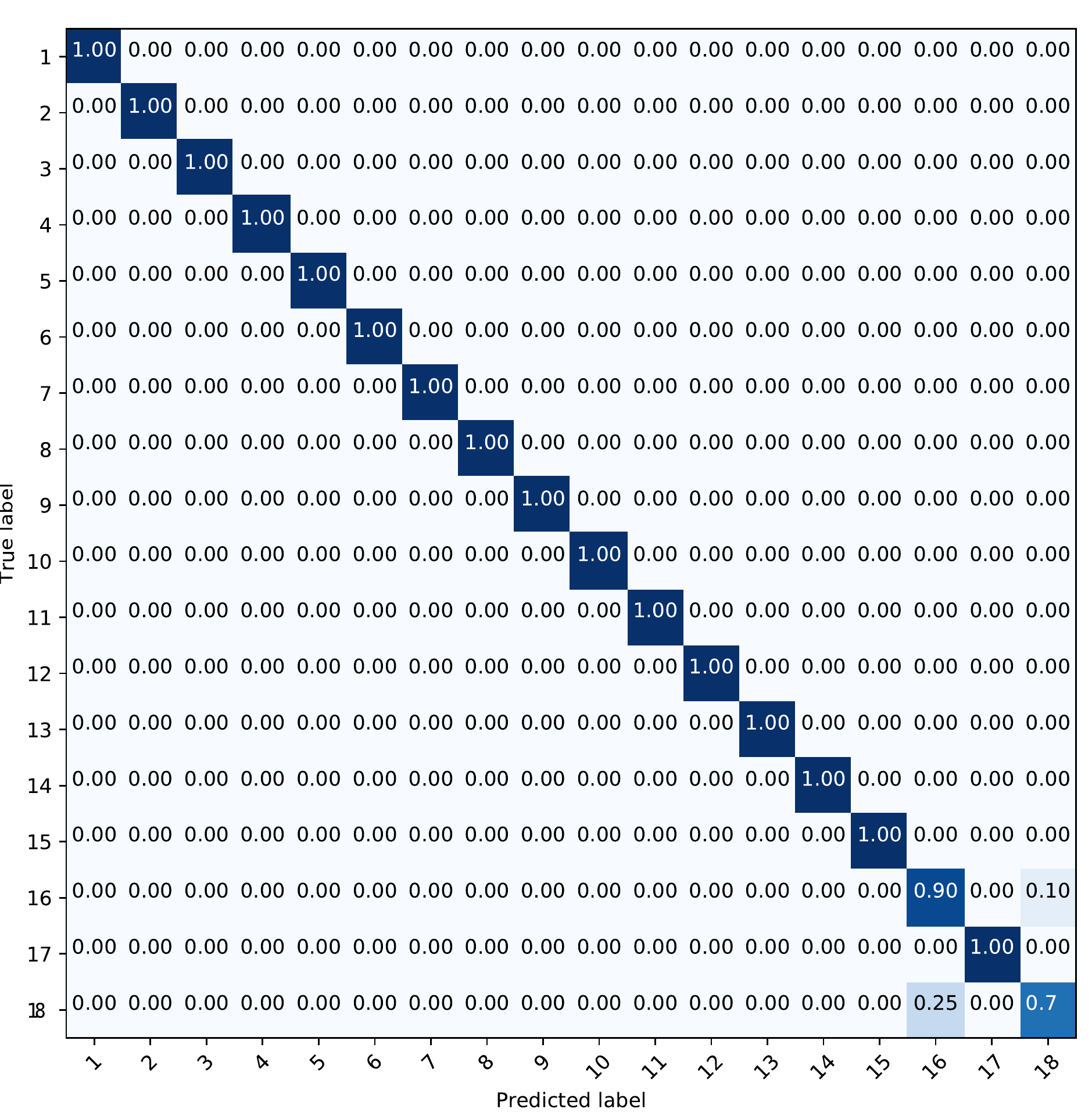}
    \caption{Confusion Matrix of our proposed RemNet trained on the augmented train set and tested on the unaltered test set of the Dresden database. The input and predicted label correspond to the Serial No. used in Table \ref{tab_dresden}.}
    \label{confo}
\end{figure}
	
After training on the augmented train set, evaluation is carried out on the unaltered test set. The results are presented in Table \ref{tab_result}. If we compare the results of Table \ref{tab_result} with that of Table \ref{tab_result_unalt}, we observe that these post-processing schemes, as a form of data-augmentation, indeed improve the performance of all the networks except that in \cite{bayar2017design}. Our proposed RemNet achieves the best accuracy of $97.59\%$ among all the models
. It is worthwhile to mention that RemNet attains 100\% accuracy on identifying 16 camera models, as shown in the corresponding confusion matrix in Fig. \ref{confo}. For the rest of the two camera models, Sony DSC-H50 and Sony DSC-W170, RemNet attains accuracy of 90\% and 75\%, respectively. The decrease in the identification accuracy for these two exact models has also been observed in \cite{tuama2016camera}. As mentioned in \cite{kirchner2015forensic}, images captured with camera models of the same manufacturer are likely to share some components which makes it harder to separate them.

\begin{figure}[!t]
	\centering
	\includegraphics[width=3.2in]{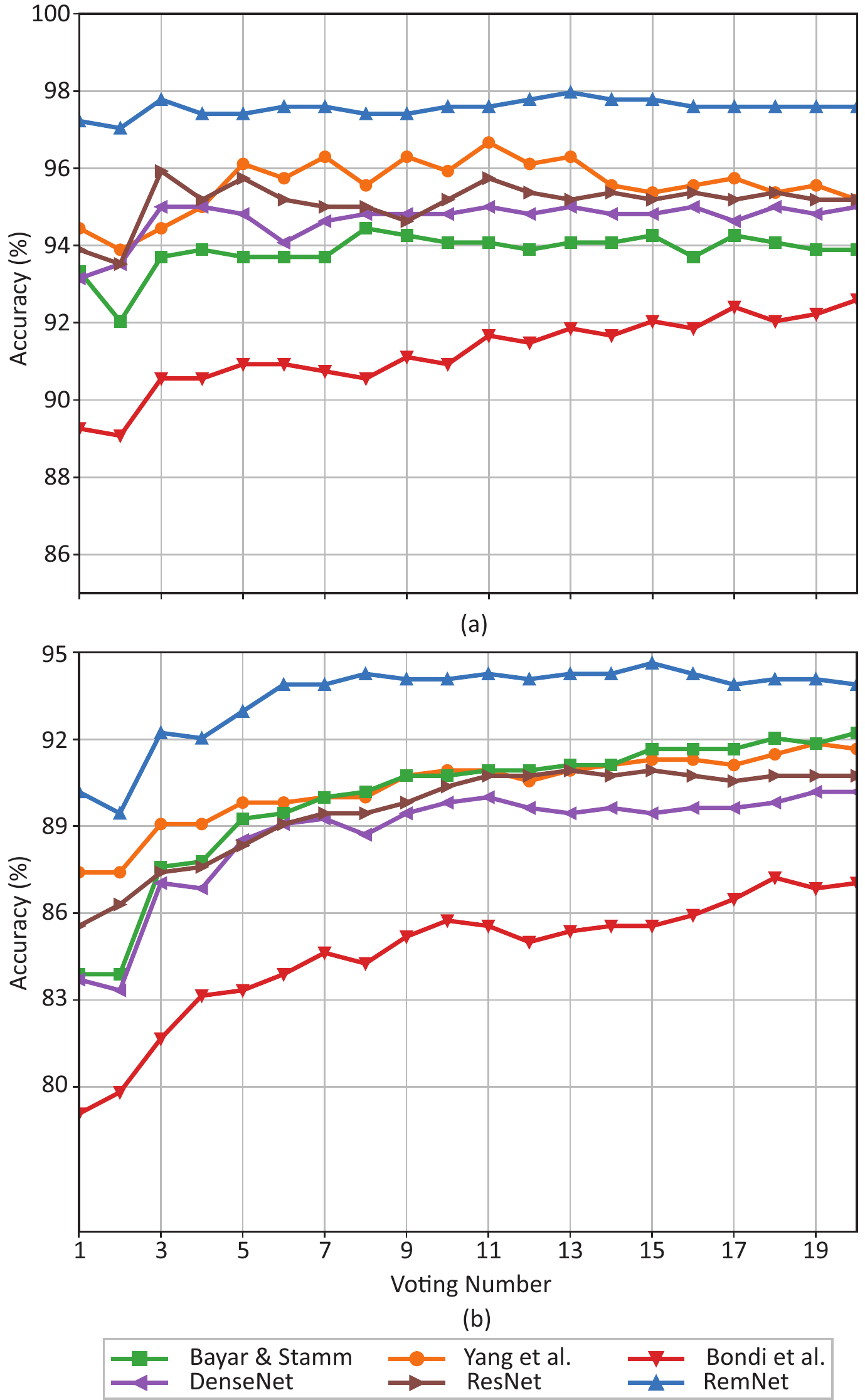}
	\caption{Results of varying voting number for (a) rich quality clusters and (b) poor quality clusters of different methods, trained on the augmented train set, for testing with the unaltered test set of the Dresden database.}
	\label{fig:patch}
\end{figure}

In Fig. \ref{fig:patch}, we observe the effect of the voting number, the number of clusters on which the prediction is made during testing, on the performance of different networks. For the rich quality clusters (see Fig. \ref{fig:patch}(a)), our network shows a somewhat steady trend, whereas the other networks show oscillatory behavior. This indicates that the performance of our network is nearly independent of the voting number of clusters, whereas an optimum voting number has to be selected for other networks. On the other hand, for prediction on poor quality clusters of an image, the accuracy gradually increases with the increment of voting number for all of the networks, as is evident from Fig. \ref{fig:patch}(b). In both of these two cases, our proposed RemNet outperforms the other networks in comparison.
	
To further ensure that the networks are not biased toward the augmented train set, we perform post-processing on test images with such factors that are not necessarily used in the augmented train and validation set. We process the test images using gamma correction with $\gamma$ = 0.5, 0.75, 1.25, and 1.5; JPEG compression quality factors (QFs) 95\%, 90\%, 85\%, and 80\%; and rescaling factor of 0.8, 0.9, 1.1, 1.2. The results of this study are presented in Table \ref{tab_manipulation}. These results show that our proposed RemNet outperforms the other networks with a significant margin in gamma correction. In rescale, the deeper models, specially ResNet \cite{he2016deep}, perform substantially better than all other networks. In JPEG compression, ResNet \cite{he2016deep} and our proposed RemNet both achieve better performances in totality.\\

\subsubsection{Significance of the Remnant Blocks}
\label{sec_significance_remnant}
In order to validate the significance of our proposed preprocessor, we train and test our proposed classifier network without the remnant blocks. We also train and test the network proposed in \cite{bondi2017first}, ResNet \cite{he2016deep}, and DenseNet \cite{densenet} together with the remnant blocks to demonstrate its generalizability to any classification network and its positive impact on their performances. All these networks are trained end-to-end on the Dresden database. It is to be mentioned that we do not perform similar experiments on  \cite{bayar2017design} and \cite{yang2017source} since these networks already consist of their own preprocessing schemes.
	
\begin{figure}[!t]
	\centering
	\includegraphics[width=3.2in]{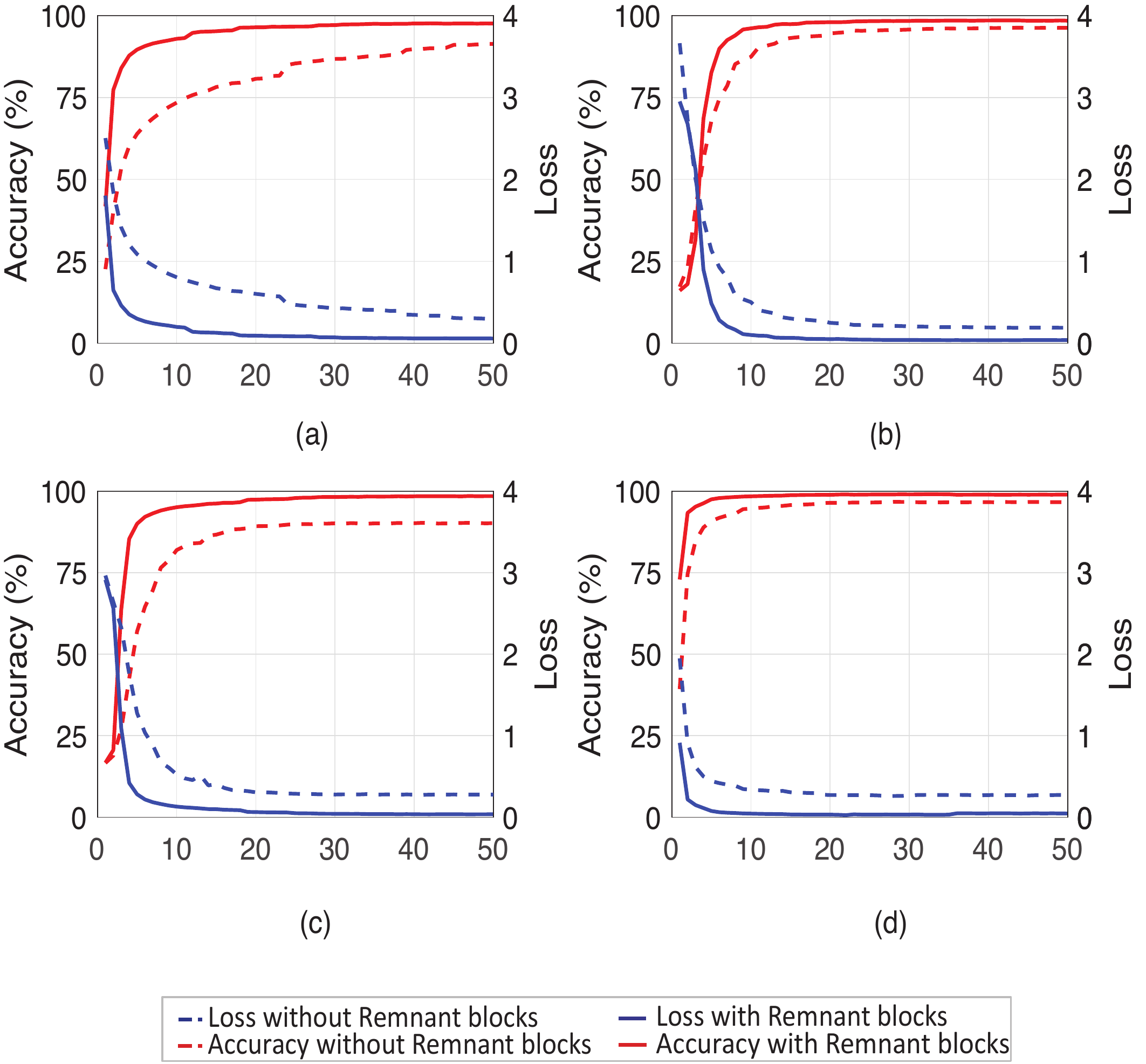}
	\caption{Training history of (a) Bondi et al., (b) DenseNet, (c) ResNet, and (d) Our Proposed Classifier, with and without remnant blocks, for training with the augmented train set of the Dresden database.}
	\label{fig_curve}
\end{figure}

The training histories of the models are presented in Fig. \ref{fig_curve}. As we can see, the addition of the remnant blocks not only improve the performances but also helps the models converge faster. The credit for these improvements can be attributed to the remnant blocks. When raw input images are fed directly to these classification networks, they are required to perform two tasks at the same time that is, to suppress the image content and to extract the required camera model fingerprints. Our proposed preprocessing scheme makes the later task easier as it suppresses the unnecessary content of the image and provides the classification block with inputs which are rich in camera model-specific features. Therefore, it becomes easier for these classification networks to identify camera models and update their weights faster during training compared to when they are trained with raw input images.

\begin{table*}[!t]
	\centering
	\begin{tabular}{C{0.9in} c c c c c c c c c c c c}
		\hline
		Manipulation & \multicolumn{4}{c}{Gamma Correction} & \multicolumn{4}{c}{JPEG Compression} & \multicolumn{4}{c}{Rescale} \\
		\hline
		Factor & 0.5 & 0.75 & 1.25 & 1.5 & 95 & 90 & 85 & 80 & 0.8 & 0.9 & 1.1 & 1.2 \\
		\hline 
		Bayar and Stamm \cite{bayar2017design} & 93.52 & 94.44 & 94.44 & 94.63 & 92.59 & 94.81 & 88.15 & 85.74 & 88.15 & 87.04 & 64.44 & 59.07 \\
		Yang et al. \cite{yang2017source} & 94.26 & 95.37 & 95.00 & 92.78 & 94.07 & 94.07 & 92.59 & 92.59 & 94.26 & 92.59 & 90.93 & 90.56 \\
		Bondi et al. \cite{bondi2017first} & 85.92 & 91.85 & 89.07 & 92.03 & 84.07 & 85.92 & 91.48 & 90.74 & 92.56 & 92.77 & 91.48 & 89.44 \\
		ResNet. \cite{he2016deep} & 91.85 & 95.18 & 92.77 & 94.81 & 93.88 & 94.82 & 95.55 & 95.00 & 95.18 & 95.18 & 95.00 & 95.18 \\
		DenseNet. \cite{densenet} & 91.66 & 95.18 & 92.03 & 94.62 & 92.77 & 92.96 & 94.26 & 94.81 & 95.00 & 94.81 & 94.44 & 94.26 \\
		Proposed Method & 96.11 & 97.22 & 96.11 & 95.56 & 97.59 & 94.81 & 92.59 & 92.78 & 95.00 & 93.33 & 92.04 & 92.41 \\
		\hline
	\end{tabular}
	\caption{Comparative results of our proposed network with different methods, trained on the augmented train set, in identifying camera models from manipulated test images of the Dresden database (Accuracy in \%)}
	\label{tab_manipulation}
\end{table*}
	
From the experimental results presented in Table \ref{tab_manip_detect}, it is clearly evident that our proposed preprocessing scheme improves the performance of all the aforementioned methods with a significant margin. The addition of our remnant blocks in cascade with these models helps them achieve substantially better performance even when they are trained with unaltered images only. Their performances further improve when they are trained with augmented data.
	
Also, in order to verify the effect of remnant blocks on the robustness of the networks trained with the augmented dataset, we further evaluate the performance of \cite{bondi2017first}, ResNet \cite{he2016deep}, and DenseNet \cite{densenet} with remnant blocks on the manipulated test dataset. The experimental results shown in Table \ref{remnant_tab_manipulation} demonstrate that with the addition of the remnant blocks, all three models have a performance gain in most of the cases and also in totality. Also, due to the adaptive nature of our preprocessing scheme and end-to-end training, the remnant blocks can learn to produce the optimum output as required by the subsequent classifier block. Such adaptive nature of our preprocessing scheme makes it a promising approach to further improve the CMI performance of the existing DNN based approaches without changing their configuration. 
	
\begin{table*}[!t]
	\centering
	\begin{tabular}{C{1in} c c c c}
		\hline
		\multirow{2}{*}{Method} & \multicolumn{2}{c}{Trained on Unaltered Train Set} & \multicolumn{2}{c}{Trained on Augmented Train Set}\\
		\cline{2-5}
		& without remnant blocks & with remnant blocks & without remnant blocks & with remnant blocks\\
		\hline 
		Bondi et al. \cite{bondi2017first}  & 90.55 & 95.92 & 92.59 & 96.29\\
		ResNet \cite{he2016deep} & 92.40 & 96.85 & 95.18 & 98.33 \\
		DenseNet \cite{densenet} & 93.33 & 96.29 & 95.01 & 98.14\\
		Proposed Classifier & 93.31 & 97.03 & 95.74 & 97.59 \\
		\hline
	\end{tabular}
	\caption{Results of different models, with and without remnant blocks, tested on the unaltered test set of the Dresden dataset (Accuracy in \%)}
	\label{tab_manip_detect}
\end{table*}

\begin{table*}[!t]
	\centering
	\begin{tabular}{C{1.2in} c c c c c c c c c c c c}
		\hline
		Manipulation & \multicolumn{4}{c}{Gamma Correction} & \multicolumn{4}{c}{JPEG Compression} & \multicolumn{4}{c}{Resize Scale} \\
		\hline
		Factor & 0.5 & 0.75 & 1.25 & 1.5 & 95 & 90 & 85 & 80 & 0.8 & 0.9 & 1.1 & 1.2 \\
		\hline
		Bondi et al. \cite{bondi2017first} & 85.92 & 91.85 & 89.07 & 92.03 & 84.07 & 85.92 & 91.48 & 90.74 & 92.56 & 92.77 & 91.48 & 89.44 \\
		Remnant-Bondi et al. & 94.07 & 95.74 & 95.37 & 95.92 & 88.88 & 89.07 & 93.52 & 92.22 & 91.66 & 91.85 & 90.00 & 88.14 \\
		ResNet. \cite{he2016deep} & 91.85 & 95.18 & 92.77 & 94.81 & 93.88 & 94.82 & 95.55 & 95.00 & 95.18 & 95.18 & 95.00 & 95.18 \\
		Remnant-ResNet & 98.33 & 98.33 & 97.59 & 97.59 & 93.33 & 93.33 & 95.18 & 95.92 & 95.37 & 95.18 & 92.40 & 95.00 \\
		DenseNet. \cite{densenet} & 91.66 & 95.18 & 92.03 & 94.62 & 92.77 & 92.96 & 94.26 & 94.81 & 95.00 & 94.81 & 94.44 & 94.26 \\
		Remnant-DenseNet. & 96.85 & 97.59 & 97.96 & 97.59 & 93.70 & 93.88 & 94.81 & 95.92 & 95.37 & 94.81 & 93.52 & 95.18 \\
		\hline
	\end{tabular}
	\caption{Comparative results of different models with and without remnant blocks, trained on the augmented train set, in identifying camera models from manipulated test images of the Dresden database (Accuracy in \%)}
	\label{remnant_tab_manipulation}
\end{table*}
	
\subsection{Results on the IEEE Signal Processing Cup 2018 Dataset}
\label{sec_result_ieee}
To test the generalizability of our approach, we have also trained and tested the aforementioned networks on the CMI Dataset provided for the IEEE Signal Processing (SP) Cup 2018 \cite{datasetsp}. The training dataset provided by the IEEE Signal Processing Society consists of images captured by 10 different camera models having 275 images for each model. Since only one device is used to capture these images for each camera model, we collect external data from multiple devices from Flickr under the creative commons license. All these images are used for training and validation purposes only. A brief summary of the dataset is given in Table \ref{tab_spdata}.

\begin{table}[!t]
	\centering
	\begin{tabular}{C{1.2in} C{0.5in} C{0.75in} }
		\hline
		\multirow{2}{*}{Camera Model} & \multicolumn{2}{c}{No. of Images}\\
		\cline{2-3} 
		& SP Cup Data & Flickr Data \\
		\hhline{||||}
		\hline 
		HTC-1-M7 & 275 & 746 \\
		iPhone-4s & 275 & 499\\
		iPhone-6 & 275 & 548\\
		LG-Nexus-5x & 275 & 405\\
		Motorola-Droid-Maxx & 275 & 549\\
		Motorola-Nexus-6 & 275	& 650\\
		Motorola-X & 275 & 344\\
		Samsung-Note3 & 275 & 274\\
		Samsung-Galaxy-S4 & 275 & 1137\\
		Sony-NEX-7 	& 275 & 557\\
		\hhline{||||}
		\hline
		Sub-Total & 2750 &5709\\
		\cline{2-3}
		Grand-Total	& \multicolumn{2}{c}{8459}\\
		\hline
	\end{tabular}
	\caption{IEEE SP Cup 2018 data and Flickr data}
	\label{tab_spdata}
\end{table}
	
The dataset described in Table \ref{tab_spdata} is split into train and validation data by a 3:1 ratio. The test dataset is provided separately, which includes 2640 images of size $512\times 512$, among which 1320 are unaltered, and the rest are augmented, i.e., resized, gamma-corrected, or JPEG compressed. All the test images are acquired with a separate device other than the ones used for capturing training and validation images.
	
The training and testing is done by following the same procedures as mentioned in the earlier experiments. This time, we train our network for 10 classes. The testing is done on the test set which contains images from completely separate devices that are used for training. Since the size of the test images is $512 \times 512$, we extract the best clusters of size $256 \times 256$ and generate result following the testing procedure mentioned previously. According to the competition rules of IEEE SP Cup 2018, the score on the test-results are calculated based on the following formula:

\begin{dmath}
	\label{eqn:5}
	\textrm{Accuracy} =  0.7\times(\textrm{Accuracy of Unaltered Images}) + \\
	0.3\times(\textrm{Accuracy of Manipulated Images})
\end{dmath}
	
Table \ref{tab_result2} summarizes the result of our model on the SP cup dataset along with comparing it with different networks. From the table, it is clear that our proposed RemNet outperforms the other networks with an accuracy of 95.11\%. This satisfactory performance is evidence of the generalizability of our approach. Among the other networks, wider (\cite{yang2017source}) and deeper (\cite{he2016deep, densenet}) networks perform comparatively better than the shallower ones. 
	
To verify the effect of remnant blocks on different networks for the IEEE SP Cup 2018 dataset, we train the networks \cite{bondi2017first}, ResNet \cite{he2016deep}, and DenseNet \cite{densenet} in cascade with remnant blocks. The experimental results are presented in Table \ref{tab_sp2}. It is clear from the table that the addition of the remnant blocks improves the performances of the aforementioned networks. Therefore, our hypothesis that the remnant blocks can improve the performance of any classification network in CMI is further verified in different datasets.  
	
\begin{table}[!b]
	\begin{tabular}{C{1.4in} C{1in}}
		\hline
		Method & Accuracy (\%)\\
		\hhline{|||}
		\hline 
		Bayar and Stamm \cite{bayar2017design} & 90.97\\
		Yang et al. \cite{yang2017source} & 94.83\\ 
		Bondi et al. \cite{bondi2017first} & 90.07\\
		ResNet \cite{he2016deep} & 93.92\\ 
		DenseNet \cite{densenet} & 93.70\\ 
		\textbf{Proposed Method} & \textbf{95.11}\\ 
		\hline
	\end{tabular}
	\caption{Accuracy of different methods on the IEEE SP Cup 2018 testing dataset}
	\label{tab_result2}
\end{table}
	
\begin{table}[!t]
	\begin{tabular}{C{1.4in} C{1in}}
		\hline
		Method & Accuracy (\%)\\
		\hhline{| | |}
		\hline
		Remnant-Bondi et al. & 92.15\\
		Remnant-ResNet & 93.98\\
		Remnant-DenseNet & 94.68\\
		\hline
	\end{tabular}
	\caption{Comparative results of different models, in cascade with remnant blocks, tested on the IEEE SP Cup 2018 testing dataset}
	\label{tab_sp2}
\end{table}
	
\begin{figure}[!b]
	\centering
	\includegraphics[width=3.2in]{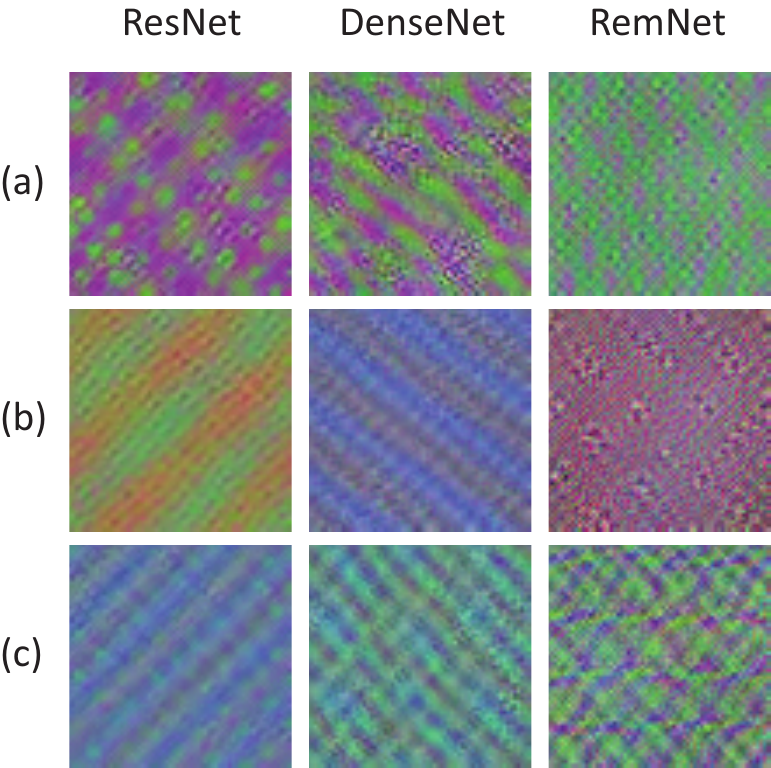}
	\caption{Visualization of input activation of (a) Canon IXUS 70, (b) CanonEX-Z150, and (c) FujiFilm FinePix J50 for different networks trained on the Dresden database.}
	\label{fig_viz}
\end{figure}

\subsection{Visualizing the Models Class Activation}
\label{sec_visualize_class_activation}
Due to a large number of parameters, the CNNs can easily get biased to the image content, rather than the intrinsic camera fingerprint. It has been, therefore, a topic of great interest among the camera-forensic experts about what type of forensic features such deep models learn for CMI. To explore this, we adopt the class activation maximization method proposed by Erhan et al \cite{activation} at the highest level of feature representation of the networks, i.e., on the output neuron to understand what type of input patterns activate the final class. The main goal of such an experiment is to observe and explore the hidden patterns present in the image that the networks have learned to extract for CMI. Due to the paper size limit, we show the generated patterns for 3 different camera models for ResNet \cite{he2016deep}, DenseNet \cite{densenet}, and our proposed network in Fig. \ref{fig_viz}. From this figure, it is evident that deep networks trained for CMI do not focus on the visible image content. The noticeable difference among the patterns of different networks can be explained by the fact that different network architecture can be thought of different transformation function to be applied to the same input, which on the other hand, may result in such a difference.

\section{Conclusion}
\label{sec_conclusion}
In this paper, a novel CNN model has been proposed for the identification of the source camera model of an image for digital image forensics. To address the CMI problem effectively, a dynamic CNN-based preprocessing block  has been placed in cascade with the shallow CNN-based classifier for enhancing the intrinsic camera model-specific fingerprints at its output by suppressing the undesired contents of the input image. Unlike the conventional fixed filter-based approaches for preprocessing in image forensics, the remnant blocks of the proposed preprocessing unit are completely data-driven. The experimental results on the Dresden and the IEEE SP Cup 2018 Camera Model Identification datasets, focusing on the unseen devices of close-set camera models and post-processed images, have demonstrated improved performance and generalizability of the proposed modular RemNet for real-world  CMI  application. Furthermore, the demonstrated ability of the remnant blocks to improve the CMI performance along with the speed of convergence of the well-known CNN-based approaches indicates that they are suitable as a general-purpose preprocessing scheme for varieties of CMI networks. In future works, we wish to explore the potential of such a preprocessing scheme in other image forensic tasks such as forgery detection and post-processing classification.\\

\begin{flushleft}
\textbf{Compliance with ethical standards}
\end{flushleft}
\vspace{1mm}

\begin{flushleft}
\textbf{Conflict of interest} All authors declare that they have no conflict of interests.
\end{flushleft}

\bibliographystyle{IEEEtran}
\bibliography{ref}

\begin{thebibliography}{10}
\providecommand{\url}[1]{#1}
\csname url@samestyle\endcsname
\providecommand{\newblock}{\relax}
\providecommand{\bibinfo}[2]{#2}
\providecommand{\BIBentrySTDinterwordspacing}{\spaceskip=0pt\relax}
\providecommand{\BIBentryALTinterwordstretchfactor}{4}
\providecommand{\BIBentryALTinterwordspacing}{\spaceskip=\fontdimen2\font plus
\BIBentryALTinterwordstretchfactor\fontdimen3\font minus
  \fontdimen4\font\relax}
\providecommand{\BIBforeignlanguage}[2]{{%
\expandafter\ifx\csname l@#1\endcsname\relax
\typeout{** WARNING: IEEEtran.bst: No hyphenation pattern has been}%
\typeout{** loaded for the language `#1'. Using the pattern for}%
\typeout{** the default language instead.}%
\else
\language=\csname l@#1\endcsname
\fi
#2}}
\providecommand{\BIBdecl}{\relax}
\BIBdecl

\bibitem{stamm2013information}
M.~C. Stamm, M.~Wu, and K.~R. Liu, ``Information forensics: An overview of the
  first decade,'' \emph{{IEEE} Access}, vol.~1, pp. 167--200, 2013.

\bibitem{san2006source}
K.~San~Choi, E.~Y. Lam, and K.~K. Wong, ``Source camera identification by jpeg
  compression statistics for image forensics,'' in \emph{TENCON IEEE
  Region}.\hskip 1em plus 0.5em minus 0.4em\relax IEEE, 2006, pp. 1--4.

\bibitem{castiglione2013experimentations}
A.~Castiglione, G.~Cattaneo, M.~Cembalo, and U.~F. Petrillo, ``Experimentations
  with source camera identification and online social networks,'' \emph{J. Amb.
  Intel. Hum. Comp}, vol.~4, no.~2, pp. 265--274, 2013.

\bibitem{kirchner2015forensic}
M.~Kirchner and T.~Gloe, ``Forensic camera model identification,'' \emph{Proc.
  WOL Handbook of Digital Forensics of Multimedia Data and Devices}, pp.
  329--374, 2015.

\bibitem{piva2013overview}
A.~Piva, ``An overview on image forensics,'' \emph{Proc. ISRN Signal Process.},
  vol. 2013, 2013.

\bibitem{farid2009image}
H.~Farid, ``Image forgery detection,'' \emph{{IEEE} Signal Process. Mag.},
  vol.~26, no.~2, pp. 16--25, 2009.

\bibitem{bayram2005source}
S.~Bayram, H.~Sencar, N.~Memon, and I.~Avcibas, ``Source camera identification
  based on cfa interpolation,'' in \emph{Proc. IEEE Int. Conf. on Image
  Process., (ICIP)}, vol.~3.\hskip 1em plus 0.5em minus 0.4em\relax IEEE, 2005,
  pp. III--69.

\bibitem{kharrazi2004blind}
M.~Kharrazi, H.~T. Sencar, and N.~Memon, ``Blind source camera
  identification,'' in \emph{Proc. IEEE Int. Conf. on Image Process., (ICIP)},
  vol.~1.\hskip 1em plus 0.5em minus 0.4em\relax IEEE, 2004, pp. 709--712.

\bibitem{gloe2012feature}
T.~Gloe, ``Feature-based forensic camera model identification,'' in \emph{LNCS
  Trans. Data Hiding and Multimed. Secur. VIII , Vol. 7228 of Lect. Notes
  Comput. Sc.}\hskip 1em plus 0.5em minus 0.4em\relax Springer, 2012, pp.
  42--62.

\bibitem{dirik2007source}
A.~E. Dirik, H.~T. Sencar, and N.~Memon, ``Source camera identification based
  on sensor dust characteristics,'' in \emph{Proc. IEEE Workshop Signal
  Process. Appl. Public Secur. Forensics}.\hskip 1em plus 0.5em minus
  0.4em\relax IEEE, 2007, pp. 1--6.

\bibitem{fridrich2006digital}
J.~Fridrich, J.~Lukas, and M.~Goljan, ``Digital camera identification from
  sensor noise,'' \emph{{IEEE} Trans. Inf. Forensics Secur.}, vol.~1, no.~2,
  pp. 205--214, 2006.

\bibitem{filler2008using}
T.~Filler, J.~Fridrich, and M.~Goljan, ``Using sensor pattern noise for camera
  model identification,'' in \emph{Proc. IEEE Int. Conf. on Image Process.,
  (ICIP)}.\hskip 1em plus 0.5em minus 0.4em\relax IEEE, 2008, pp. 1296--1299.

\bibitem{thai2014camera}
T.~H. Thai, R.~Cogranne, and F.~Retraint, ``Camera model identification based
  on the heteroscedastic noise model,'' \emph{{IEEE} Trans. Image Process.},
  vol.~23, no.~1, pp. 250--263, 2014.

\bibitem{lukas2006digital}
J.~Lukas, J.~Fridrich, and M.~Goljan, ``Digital camera identification from
  sensor pattern noise,'' \emph{{IEEE} Trans. Inf. Forensics Secur.}, vol.~1,
  no.~2, pp. 205--214, 2006.

\bibitem{cao2009accurate}
H.~Cao and A.~C. Kot, ``Accurate detection of demosaicing regularity for
  digital image forensics,'' \emph{{IEEE} Trans. Inf. Forensics Secur.},
  vol.~4, no.~4, pp. 899--910, 2009.

\bibitem{swaminathan2007nonintrusive}
A.~Swaminathan, M.~Wu, and K.~R. Liu, ``Nonintrusive component forensics of
  visual sensors using output images,'' \emph{{IEEE} Trans. Inf. Forensics
  Secur.}, vol.~2, no.~1, pp. 91--106, 2007.

\bibitem{chen2015camera}
C.~Chen and M.~C. Stamm, ``Camera model identification framework using an
  ensemble of demosaicing features,'' in \emph{Proc. IEEE Int. Works. Infor.
  (WIFS)}.\hskip 1em plus 0.5em minus 0.4em\relax IEEE, 2015, pp. 1--6.

\bibitem{marra2017study}
F.~Marra, G.~Poggi, C.~Sansone, and L.~Verdoliva, ``A study of co-occurrence
  based local features for camera model identification,'' \emph{Multimedia
  Tools and Applications}, vol.~76, no.~4, pp. 4765--4781, 2017.

\bibitem{chen2015median}
J.~Chen, X.~Kang, Y.~Liu, and Z.~J. Wang, ``Median filtering forensics based on
  convolutional neural networks,'' \emph{{IEEE} Signal Process. Lett.},
  vol.~22, no.~11, pp. 1849--1853, 2015.

\bibitem{tuama2016camera}
A.~Tuama, F.~Comby, and M.~Chaumont, ``Camera model identification with the use
  of deep convolutional neural networks,'' in \emph{Proc. IEEE Int. Workshop on
  inf. Forensics and Secur. (WIFS)}.\hskip 1em plus 0.5em minus 0.4em\relax
  IEEE, 2016, pp. 1--6.

\bibitem{bayar2017design}
B.~Bayar and M.~C. Stamm, ``Design principles of convolutional neural networks
  for multimedia forensics,'' \emph{Electronic Imaging}, vol. 2017, no.~7, pp.
  77--86, 2017.

\bibitem{bayar2016deep}
------, ``A deep learning approach to universal image manipulation detection
  using a new convolutional layer,'' in \emph{Proc. 4th-ACM Workshop on inf.
  Hiding and Multimedia Secur.}\hskip 1em plus 0.5em minus 0.4em\relax ACM,
  2016, pp. 5--10.

\bibitem{yang2017source}
P.~Yang, W.~Zhao, R.~Ni, and Y.~Zhao, ``Source camera identification based on
  content-adaptive fusion network,'' \emph{Pattern Recog. Lett.}, vol. 119, pp.
  195--204, 2019.

\bibitem{bondi2017first}
L.~Bondi, L.~Baroffio, D.~G{\"u}era, P.~Bestagini, E.~J. Delp, and S.~Tubaro,
  ``First steps toward camera model identification with convolutional neural
  networks,'' \emph{{IEEE} Signal Process. Lett.}, vol.~24, no.~3, pp.
  259--263, 2017.

\bibitem{yao2018robust}
H.~Yao, T.~Qiao, M.~Xu, and N.~Zheng, ``Robust multi-classifier for camera
  model identification based on convolution neural network,'' \emph{{IEEE}
  Access}, vol.~6, pp. 24\,973--24\,982, 2018.

\bibitem{rafi2019application}
A.~M. Rafi, U.~Kamal, R.~Hoque, A.~Abrar, S.~Das, R.~Lagani{\`e}re, and M.~K.
  Hasan, ``Application of densenet in camera model identification and
  post-processing detection.'' in \emph{CVPR Workshops}, 2019, pp. 19--28.

\bibitem{huang2017densely}
G.~Huang, Z.~Liu, K.~Q. Weinberger, and L.~van~der Maaten, ``Densely connected
  convolutional networks,'' in \emph{Proc. IEEE Conf. Comput. Vision Pattern
  Recognit. (CVPR)}, vol.~1, no.~2, 2017, p.~3.

\bibitem{li2008detecting}
B.~Li, Y.~Q. Shi, and J.~Huang, ``Detecting doubly compressed jpeg images by
  using mode based first digit features,'' in \emph{Proc. IEEE 10th Workshop on
  Multimedia Signal Process.}\hskip 1em plus 0.5em minus 0.4em\relax IEEE,
  2008, pp. 730--735.

\bibitem{stamm2010forensic}
M.~C. Stamm and K.~R. Liu, ``Forensic detection of image manipulation using
  statistical intrinsic fingerprints,'' \emph{{IEEE} Trans. Inf. Forensics
  Secur.}, vol.~5, no.~3, pp. 492--506, 2010.

\bibitem{kee2011digital}
E.~Kee, M.~K. Johnson, and H.~Farid, ``Digital image authentication from jpeg
  headers,'' \emph{{IEEE} Trans. Inf. Forensics Secur.}, vol.~6, no.~3, pp.
  1066--1075, 2011.

\bibitem{srivastava2015highway}
R.~K. Srivastava, K.~Greff, and J.~Schmidhuber, ``Highway networks,''
  \emph{arXiv preprint arXiv:1505.00387}, 2015.

\bibitem{he2016deep}
K.~He, X.~Zhang, S.~Ren, and J.~Sun, ``Deep residual learning for image
  recognition,'' in \emph{Proc. IEEE Conf. Comput. Vision Pattern Recognit.
  (CVPR)}, 2016, pp. 770--778.

\bibitem{greff2016highway}
K.~Greff, R.~K. Srivastava, and J.~Schmidhuber, ``Highway and residual networks
  learn unrolled iterative estimation,'' \emph{arXiv preprint
  arXiv:1612.07771}, 2016.

\bibitem{prnu}
Ashref, Lawgaly, Fouad, and Khelifi, ``Sensor pattern noise estimation based on
  improved locally adaptive dct filtering and weighted averaging for source
  camera identification and verification,'' \emph{{IEEE} Trans. Inf. Forensics
  Secur.}, vol.~12, 2017.

\bibitem{springenberg2014striving}
J.~T. Springenberg, A.~Dosovitskiy, T.~Brox, and M.~Riedmiller, ``Striving for
  simplicity: The all convolutional net,'' \emph{arXiv preprint
  arXiv:1412.6806}, 2014.

\bibitem{krizhevsky2012imagenet}
A.~Krizhevsky, I.~Sutskever, and G.~E. Hinton, ``Imagenet classification with
  deep convolutional neural networks,'' in \emph{Adv. in Neural Inf. Process.
  Systems}, 2012, pp. 1097--1105.

\bibitem{simonyan2014very}
K.~Simonyan and A.~Zisserman, ``Very deep convolutional networks for
  large-scale image recognition,'' \emph{arXiv preprint arXiv:1409.1556}, 2014.

\bibitem{dresden}
T.~Gloe and R.~B\"ohme, ``The dresden image database for benchmarking digital
  image forensics.'' \emph{J. Digital Forensic Practice}, vol.~3, pp. 150--159,
  01 2010.

\bibitem{glorot2010understanding}
X.~Glorot and Y.~Bengio, ``Understanding the difficulty of training deep
  feedforward neural networks,'' in \emph{Proc. AISTATS}, 2010, pp. 249--256.

\bibitem{kingma2014adam}
D.~P. Kingma and J.~Ba, ``Adam: A method for stochastic optimization,''
  \emph{arXiv preprint arXiv:1412.6980}, 2014.

\bibitem{Chen_densenet}
Y.~Chen, X.~Kang, Z.~J. Wang, and Q.~Zhang, ``Densely connected convolutional
  neural network for multi-purpose image forensics under anti-forensic
  attacks,'' in \emph{Proc. 6th ACM Workshop Inf. Hiding Multimedia
  Secur.}\hskip 1em plus 0.5em minus 0.4em\relax New York, NY, USA: ACM, 2018,
  pp. 91--96.

\bibitem{barni-jpeg-cnn}
M.~{Barni}, A.~{Costanzo}, E.~{Nowroozi}, and B.~{Tondi}, ``Cnn-based detection
  of generic contrast adjustment with jpeg post-processing,'' in \emph{Proc.
  IEEE Int. Conf. on Image Process. (ICIP)}, Oct 2018, pp. 3803--3807.

\bibitem{DL_cmi}
F.~J. Boroumand, Mehdi, ``Deep learning for detecting processing history of
  images,'' \emph{Electronic Imaging}, 2018.

\bibitem{densenet}
G.~{Huang}, Z.~{Liu}, L.~v.~d. {Maaten}, and K.~Q. {Weinberger}, ``Densely
  connected convolutional networks,'' in \emph{Proc. IEEE Conf. Comput. Vision
  Pattern Recognit. (CVPR)}, July 2017, pp. 2261--2269.

\bibitem{wong2016understanding}
S.~C. Wong, A.~Gatt, V.~Stamatescu, and M.~D. McDonnell, ``Understanding data
  augmentation for classification: when to warp?'' in \emph{Int. Conf. Digit.
  Image Comput.: Tech. and Appl. (DICTA)}.\hskip 1em plus 0.5em minus
  0.4em\relax IEEE, 2016, pp. 1--6.

\bibitem{datasetsp}
M.~Stamm, P.~Bestagini, L.~Marcenaro, and P.~Campisi, ``Forensic camera model
  identification: Highlights from the ieee signal processing cup 2018 student
  competition [sp competitions],'' \emph{IEEE Signal Process. Mag.}, vol.~35,
  no.~5, pp. 168--174, 2018.

\bibitem{activation}
D.~Erhan, Y.~Bengio, A.~Courville, and P.~Vincent, ``Visualizing higher-layer
  features of a deep network,'' \emph{University of Montreal}, vol. 1341, p.~1,
  2009.

\end{thebibliography}

\end{document}